\documentclass[12pt]{article}
\usepackage{amsmath,verbatim}
\usepackage{amsfonts}
\usepackage{graphicx}
\usepackage{enumerate}
\usepackage{floatrow}
\usepackage{multirow}
\usepackage{setspace}
\newfloatcommand{capbtabbox}{table}[][\FBwidth]
\usepackage{natbib}
\usepackage{url} 
\usepackage{xcolor}
\usepackage{hyperref}

\newcommand{\xo}{X^{(1)}}
\newcommand{\xt}{X^{(2)}}
\newcommand{\yo}{Y^{(1)}}
\newcommand{\yt}{Y^{(2)}}
\newcommand{\xM}{X^{(K)}}

\newcommand{\blind}{1}

\newtheorem{example}{Example}
\newtheorem{remark}{Remark}
\newtheorem{definition}{Definition}

\newtheorem{proposition}{Proposition}

\begin{document}

\def\spacingset#1{\renewcommand{\baselinestretch}%
{#1}\small\normalsize} \spacingset{1}


\if1\blind{
  \title{\bf Discussion of ``Data fission: splitting a single data point"}
  \author{Anna Neufeld \\
  Department of Mathematics and Statistics, Williams College \\ \hspace{.2cm} \\
  Ameer Dharamshi \\ 
  Department of Biostatistics, University of Washington \\ \\
  Lucy L. Gao \\ 
  Department of Statistics, University of British Columbia \\ \\
  Daniela Witten \\ 
  Departments of Statistics and Biostatistics, University of Washington \\ \\
  Jacob Bien \\
 Department of Data Sciences and Operations, University of Southern California
 }
  \maketitle
} \fi
\bigskip
\begin{abstract}
\cite{leiner2023data} introduce an important generalization of sample splitting, which they call \emph{data fission}. They consider two cases of data fission: \emph{P1 fission} and \emph{P2 fission}. While P1 fission is extremely useful and easy to use, \cite{leiner2023data} provide P1 fission operations only for the Gaussian and the Poisson distributions. They provide little guidance on how to apply P2 fission operations in practice, leaving the reader unsure of how to apply data fission outside of the Gaussian and Poisson settings. In this discussion, we describe how our own work provides P1 fission operations  in a wide variety of families and offers insight into when P1 fission is possible. We also provide guidance on how to actually apply P2 fission in practice, with a special focus on logistic regression. Finally, we interpret P2 fission as a remedy for distributional misspecification when carrying out P1 fission operations.
\end{abstract}
 
\newpage
\spacingset{1.8} 

\section{Introduction}

Sample splitting, in which some of the observations in a data sample are used as a training set for model fitting or model selection, and the remaining observations are used as a test set for validation or inference, is a fundamental tool in data analysis. We applaud \cite{leiner2023data} for their important paper, which introduces  \emph{data fission}, a broad generalization of sample splitting.  

Data fission refers to any operation that splits a random variable $X \sim P_{\theta}$ into two pieces, which we denote here as $\xo$ and $\xt$, 
such that three properties hold: (1) there exists a deterministic function $T(\cdot)$ such that $X = T\left(\xo, \xt\right)$; (2) one cannot reconstruct $X$ from $\xo$ alone;  and (3) the marginal distribution of $\xo$ and the conditional distribution of $\xt \mid \xo$ are known, up to the unknown parameter $\theta$. \cite{leiner2023data} use the term {\em P1 fission} for the special case in which $\xo$ and $\xt$ are independent, and  \emph{P2 fission} for when independence does not necessarily hold.\footnote{For clarity, we will use ``P2 fission'' to refer specifically to fission that is not P1 fission, i.e. the case in which $\xo$ and $\xt$ are not independent.  Otherwise,  ``fission'' and ``P2 fission'' would be synonymous.}
If $X$ is a vector of independent observations, then sample splitting is an instance of P1 fission. However, data fission extends far beyond sample splitting, even to settings where we observe only a single scalar random variable $X$ that we wish to partition into two pieces. Such a generalization is useful in settings where sample splitting is unsatisfying or inapplicable, such as those discussed in \cite{leiner2023data, neufeld2024data}, and \cite{dharamshi2024generalized}. 

The main contributions of our discussion are as follows: 
\begin{enumerate}
    \item In Section~\ref{section_independence}, we argue that P1 fission is preferable to P2 fission when both are available, both due to its simplicity but also due to a new information inequality. We also show that P1 fission is available for many more distributions than the two noted by \citet{leiner2023data}.
      \item In Section \ref{sec_logistic}, we revisit logistic regression. While this is a setting where P1 fission  is impossible, we provide a major improvement on \cite{leiner2023data}'s treatment of logistic regression using P2 fission: we  conduct inference on the parameters of interest rather than on targets of convenience.
      \item In Section \ref{section_p2}, we show that P2 fission can sometimes be interpreted as P1 fission under model misspecification. This suggests room for improvement in the way that previous authors have handled P1 fission under misspecification, and suggests an avenue for developing new P2 fission operations.
\end{enumerate}

\section{The case for P1 fission over P2 fission}
\label{section_independence}

While \cite{leiner2023data} provide a very large number of P2 fission examples, they provide  only two examples of P1 fission: the Gaussian location family and the Poisson. Furthermore, for these two examples,  they also provide P2 fission alternatives.

This may give the reader the impression that P1 fission is rarely available, and that P2 fission is as effective as P1 fission when both are possible.
In Section \ref{sec:information}, we argue that P1 fission is preferable to P2 fission when both are available.  In Section \ref{section_systematic}, we describe our recent work that establishes that P1 fission is in fact widely available. 

\subsection{P1 fission is preferable over P2 fission when both are available}
\label{sec:information}
P1 fission decomposes $X$ into independent parts, while P2 fission yields dependent parts.  At an intuitive level, having independent parts may seem preferable in the name of simplicity.  Indeed, in the two examples where \cite{leiner2023data} have both P1 and P2 strategies available, they choose to only work with the P1 strategies for numerical experiments (and a P2 strategy that they carry out in the supplement involves more technical machinery such as quasi-likelihood).  However, putting aside convenience, one may wonder whether there is a {\em statistical} advantage to the independence offered by P1 fission. The following proposition answers this question in the affirmative. For ease of presentation, we focus on a scalar-valued $\theta$,  though a similar result holds for general $\theta$.

\begin{proposition}[Fisher information allocation of P1 vs. P2]\label{prop1}
Suppose $X\sim P_\theta$, where $P_\theta$ belongs to a family $\mathcal{P}$, parameterized by $\theta\in\mathbb R$, for which both P1 and P2 fission operations are available. Let $(\xo,\xt)$ and $(\tilde{X}^{(1)},\tilde{X}^{(2)})$ denote the results of P1 fission and P2 fission, respectively.  Suppose that the Fisher informations $I_X(\theta)$, $I_{\xo}(\theta)$, $I_{\xt}(\theta)$, and $I_{\tilde{X}^{(1)}}(\theta)$ all exist, as well as  $I_{\tilde{X}^{(2)}\mid \tilde{X}^{(1)}}(\theta)$, which denotes the Fisher information contained in $\tilde{X}^{(2)}$ conditional on $\tilde{X}^{(1)}$ (this is a random quantity whose value depends on $\tilde{X}^{(1)}$). 
If the two fission strategies allocate equal information to the training set, i.e. $I_{\xo}(\theta)=I_{\tilde{X}^{(1)}}(\theta)$, then on the test sets,
\begin{equation}
\mathbb E_{\tilde{X}^{(1)}} \left[\left( I_{\tilde{X}^{(2)}\mid\tilde{X}^{(1)}}(\theta) \right) ^{-1}\right]\ge \left( I_{\xt}(\theta) \right)^{-1}.
\label{eq:prop1}
\end{equation}
\end{proposition}

This result, which we prove in Appendix~\ref{app:info-proof}, is directly inspired by \cite{rasines2023splitting}'s Proposition 1 (restated in \cite{leiner2023data} as ``Fact 1").\footnote{As noted in \cite{rasines2023splitting}, ``optimality of the Fisher information is commonly measured through summary statistics of its inverse."} They use this to argue that the deterministic allocation of Fisher information of P1 fission is more efficient than sample splitting. Our result, which applies the same logic, establishes that P1 fission is at least as efficient as any P2 fission strategy. The next example illustrates that the efficiency loss can be extreme.

\begin{remark}[Information allocation for Poisson]
\label{remark_poisson_p2}
Let $X \sim \mathrm{Poisson}(\theta)$. If we apply the P1 fission operation from \citet{leiner2023data} to $X$ with tuning parameter $\epsilon$, we have that $X^{(2)} \sim \mathrm{Poisson}\left( (1-\epsilon) \theta \right)$ and $\left(I_{X^{(2)}}(\theta)\right)^{-1} = \frac{\theta}{1-\epsilon}$. If we apply the P2 fission operation from the supplementary materials of \citet{leiner2023data} to $X$ with tuning parameter $\tau$, we have that
$\tilde{X}^{(2)} \mid \tilde{X}^{(1)}\sim \mathrm{Binomial}\left(\tilde{X}^{(1)}, \frac{\theta}{\theta+\tau}\right)$,
so that 
$I_{\tilde{X}^{(2)}\mid\tilde{X}^{(1)}}(\theta)= \frac{\tau \tilde{X}^{(1)}}{\theta (\theta+\tau)^2}$.
Since $\tilde X^{(1)} \sim \mathrm{Poisson}(\theta + \tau)$, there is a non-zero probability that $\tilde{X}^{(1)}=0$. Thus, 
$\mathbb{E}_{\tilde X^{(1)}}\left[ \left(I_{\tilde{X}^{(2)}\mid\tilde{X}^{(1)}}(\theta)\right)^{-1}\right] = \infty$, i.e. confidence intervals for $\theta$ based on P2 fission would have infinite expected width while $\left(I_{X^{(2)}}(\theta)\right)^{-1} = \frac{\theta}{1-\epsilon}$ is finite.
\end{remark}
Thus, in addition to the simplicity that comes with independence, there is also a statistical justification for preferring P1 fission when it is available. 

\subsection{A systematic recipe for P1 fission}
\label{section_systematic}

In light of the advantages of P1 fission, it is natural to wonder whether P1 fission is possible beyond the normal-location and Poisson families, and what are 
 the underlying principles that determine whether P1 fission is possible in a particular family.

In our own work, we have answered these questions, showing that P1 fission is in fact widely available \citep{neufeld2024data, dharamshi2024generalized}. In Table~\ref{table_maintable}, we show that in many of the families for which \cite{leiner2023data} provide P2 fission operations  (either through their conjugate prior strategy or otherwise), P1 fission operations are also available.

Furthermore, our work  elucidates the general principles for when P1 fission is possible, and how a P1 fission operation might be constructed.  We first define \emph{data thinning}, which is a $K$-fold generalization of P1 fission (which itself is equivalent to the ``$(U,V)$-decomposition" defined by \citealt{rasines2023splitting}). 

 \begin{definition}[Data thinning, \cite{dharamshi2024generalized}]
\label{def_thinning}
Consider a family of distributions $\mathcal{P} = \{ P_\theta : \theta \in \Theta \}$. We say that this family is \emph{thinned} by a function $T\left(\cdot\right)$ if there exists a distribution $G_t$, not depending on $\theta$, such that when we draw $\left(\xo, \xt, \ldots, \xM \right) \mid X \sim G_X$, it holds that: 
\begin{enumerate}
\item $X = T\left(\xo, \ldots, \xM \right)$. 
\item $\left(\xo, \ldots, \xM \right)$ are mutually independent with  distributions that depend on $\theta$. 
\end{enumerate}
\end{definition}
  
With this definition in place, \cite{dharamshi2024generalized}'s Theorem 1 shows that sufficiency is a key ingredient in making data thinning possible.
As seen in Table~\ref{table_maintable}, this framework allows us to define P1 fission operations (equivalently, thinning operations) for a wide variety of settings, both within exponential families and beyond. 

\begin{table}
\scriptsize
\begin{spacing}{1.5}
\begin{tabular}{|c || c | c | c ||}
\hline
& Distribution $P_{\theta}$ & P1 & P2 \\
\hline
\hline
\multirow{10}{*}{
\begin{tabular}{c}
P1 and P2 available,  \\
P1 preferable. 
\end{tabular}
}
& $N(\theta, \sigma^2)$ & \cite{rasines2023splitting}  &  \multirow{10}{*}{\shortstack[c]{\vdots\\\cite{leiner2023data}\\\vdots}}   \\
& $\mathrm{N}_p(\boldsymbol\theta
, \boldsymbol\Sigma)$ &\cite{rasines2023splitting} &  \\
& $\mathrm{Poisson}(\theta)$ &  \cite{leiner2023data} & \\
\cline{2-3}
& $\mathrm{NegBin}(r, \theta)$ & \multirow{4}{*}{\shortstack[c]{\vdots\\\cite{neufeld2024data}\\\vdots}} &   \\
& $\mathrm{Binomial}(r,\theta) $ & &\\
& $\mathrm{Exp}(\theta)$ &  & \\
& $\mathrm{Gamma}(\alpha, \theta)$ & & \\
 \cline{2-3} 
& $\mathrm{Gamma}(\theta, \beta)$ &  \multirow{3}{*}{\shortstack[c]{\vdots\\\cite{dharamshi2024generalized}\\\vdots}} & \\
& $\mathrm{Dirichlet(\bold{\theta}, \phi)}$ & & \\
& Exponential family & & \\
\hline 
\multirow{6}{*}{
\begin{tabular}{c}
P1 available, \\
P2 not yet considered. 
\end{tabular}
}
& $\mathrm{Multinomial}_p(r, \boldsymbol\theta)$ & \cite{neufeld2024data} &  \\
\cline{2-3} 
& $\mathrm{Beta}(\theta,\beta)$ & \multirow{8}{*}{\shortstack[c]{\vdots\\\cite{dharamshi2024generalized}\\\vdots}} & \multirow{8}{*}{\shortstack[c]{\vdots\\Not yet considered\\\vdots}} \\
& $\mathrm{Beta}(\alpha, \theta)$ & & \\
 & $\mathrm{Weibull}(\theta, \gamma)$ &  &\\
& $\mathrm{Pareto}(\gamma, \theta)$ &  &\\
& $\mathrm{N}(\mu, \theta)$ &  & \\
& $\mathrm{Unif}(0,\theta)$ & & \\
& $\theta \cdot\mathrm{Beta}(\alpha, 1)$ & &\\
& $\theta + \mathrm{Exp}(\lambda)$ & & \\
\hline 
\multirow{3}{*}{
\begin{tabular}{c}
P2 available, \\
P1 impossible or \\
not yet considered. 
\end{tabular}
}
& $\mathrm{Bernoulli}(\theta)$ & Impossible (\cite{dharamshi2024generalized}) &  \multirow{5}{*}{\shortstack[c]{\vdots\\\cite{leiner2023data}\\\vdots}} \\ 
& $\mathrm{Categorical}(\theta)$ & Impossible (\cite{dharamshi2024generalized})&  \\
& $\mathrm{N}_1(\theta_1, \theta_2)$ &  Impossible (\cite{dharamshi2024decomposing}) & \\
& $\mathrm{Gamma}(\theta_1, \theta_2)$ & Not yet considered& \\
& $\mathrm{NegBin}(\theta_1, \theta_2)$ & Not yet considered& \\
& $\mathrm{Binomial}(\theta_1, \theta_2)$ & Not yet considered& \\
\hline 
\end{tabular}
\end{spacing}
\caption{{\em\footnotesize The parameter $\theta$ (likewise $\theta_1$ and $\theta_2$) is always unknown (i.e. the fission operation must not rely on it) while all other parameters are known. For each family, we include a citation to the authors who, to the best of our knowledge, first proposed the decomposition as a general alternative to sample splitting. We have omitted citations to authors who proposed the decompositions for specific tasks. For example, the P1 decomposition of the Poisson follows from a much-used thinning property (see e.g. \cite{chen2021estimating} and \cite{sarkar2021separating}'s use 
for specific tasks related to model validation, or \cite{neufeld2024inference}'s use for a specific task related to inference). Similarly, \cite{tian2018selective} and \cite{tian2020prediction} use the $N(\theta,\sigma^2)$ decomposition for specific tasks. Finally, \cite{joe1996time} use the natural exponential family P1 decompositions as generative models for time series. Finally, we note that this table is not exhaustive; some of the distributions considered in \cite{dharamshi2024decomposing} that are omitted here for brevity.}
}
\label{table_maintable}
\end{table}

In light of Section \ref{sec:information}, Table~\ref{table_maintable} shows that, for many of the distributions considered in \cite{leiner2023data}, P1 fission operations are available and preferable. This may leave one to wonder whether P2 fission should ever be used in practice. As highlighted in the table, \cite{dharamshi2024generalized} and \cite{dharamshi2024decomposing} establish that there are situations in which P1 fission is unavailable or impossible; in these cases, P2 fission is indeed important in practice.  The remainder of our discussion focuses on these cases.

\section{Improving P2 fission for logistic regression}
\label{sec_logistic}

In their main text, \citet{leiner2023data} do not offer concrete guidance on applying P2 fission in practical contexts. However, Section E.4 of their supplementary materials does provide one case study of applying P2 fission to a concrete problem, namely logistic regression.  This is an important example for P2 fission since, as noted in Table~\ref{table_maintable}, \cite{dharamshi2024generalized} prove that  P1 fission is impossible in the Bernoulli family.

In Section~\ref{sec:leiner-logistic}, we review their proposal and point out a major shortcoming: they are not able to conduct inference on the parameters of interest. Then, in Section \ref{sec:our-logistic}, we show that we can improve their example, establishing more persuasively the practical usefulness of P2 fission. 

\subsection{ \cite{leiner2023data}'s proposal for logistic regression}
\label{sec:leiner-logistic}

In Section A of their supplementary materials,  \cite{leiner2023data} describe a P2 fission of the Bernoulli distribution, which for reference we restate here.

\begin{example}[P2 fission of a Bernoulli \citep{leiner2023data}] \label{ex:bernoulli}
We observe $Y \sim \mathrm{Bernoulli}(\theta)$. For a tuning parameter $\epsilon$, sample $Z \sim \mathrm{Bernoulli}(\epsilon)$ and then let $\yo = \left(1-Z\right) Y + Z\left(1-Y\right)$ and $\yt=Y$, which  yields
\begin{align}
\yo &\sim \mathrm{Bernoulli}(\theta + \epsilon - 2\theta \epsilon), \nonumber \\
\yt \mid \yo  &\sim 
\label{eq_bernoulli_leiner_conditional}
\mathrm{Bernoulli}\left(\frac{\theta}{\theta + (1-\theta)(\frac{\epsilon}{1-\epsilon})^{2 \yo-1}}\right). 
\end{align}
\end{example}

In Section E.4 of the supplementary materials, \cite{leiner2023data} apply this decomposition to logistic regression. We briefly describe this example below.

\begin{example}[Inference for logistic regression using P2 fission of  a Bernoulli]
\label{ex_leiner_bernoulli}
Let $X \in \mathbb{R}^{n \times p}$ be a (fixed) matrix of covariates with rows denoted $x_i$, and let $Y \in \{0,1\}^n$ be a vector of binary responses.  Assume that 
\begin{equation}
    Y_i \sim \mathrm{Bernoulli}(\theta_i),
\text{  where  } \theta_i = \frac{\exp(\beta^T x_i)}{1 + \exp(\beta^T x_i)},
\label{eq_bernoulli_model}
\end{equation}
and  $\beta \in \mathbb{R}^p$ is sparse. \cite{leiner2023data} suggest the following workflow for  inference on the non-zero coefficients. 
\begin{enumerate}
\item For $i=1,\ldots, n$, apply the decomposition in Example~\ref{ex:bernoulli}  to split $Y_i$ into $\yo_i$ and $\yt_i$.
\item Select variables $\mathcal{S} \subseteq \{1,\ldots,p\}$ by running standard logistic lasso software (with cross-validation to select the penalty parameter) on $\left\{ (x_1, \yo_1), \ldots, (x_n, \yo_n ) \right\}$.
\item Fit a standard logistic GLM  on $\left\{ (x_{1S}, \yt_1 ), \ldots, (x_{nS}, \yt_n) \right\}$, where $x_{iS}$ denotes the $i$th observation subset to the variables in $S$. This is a misspecified model, since  \eqref{eq_bernoulli_leiner_conditional} implies that the log odds of the mean of $Y^{(2)}$ is not linear in the covariates. Thus, use sandwich standard errors for inference to obtain valid confidence intervals for the parameters that minimize the KL-divergence between the ``working model" (assumed by the use of a standard logistic GLM) and the conditional distribution in \eqref{eq_bernoulli_leiner_conditional}.  
\end{enumerate} 
\end{example}
We note that the parameters that are targeted in Step 3 of their approach are \emph{not} the same parameters that appear in the original model \eqref{eq_bernoulli_model}.  This is a serious practical limitation of the approach, as we demonstrate here. We
 generate $2,000$ datasets with $n=500$ and $p=50$ where $\beta_0=0.6$ and $\beta_j=0$ for $j=1,\ldots,p$. Note that $Y_i \sim \mathrm{Bernoulli}(0.6)$ for $i=1,\ldots,n$, and so none of the covariates  contribute to the data generating mechanism. For each dataset, we carry out the three-step process described in Example~\ref{ex_leiner_bernoulli}, which follows Section E.4 of \cite{leiner2023data}. Figure~\ref{fig_globalnull} shows that the p-values for the selected variables do not follow a uniform distribution, even though $\beta_1=\ldots=\beta_p=0$. Thus, Type 1 error of the selected variables is not controlled. 

 As this is the only example of P2 fission given in \cite{leiner2023data},  it might appear that P2 fission never lends itself to inference on the parameters of interest. However, we show in the next section that this is not the case. With care, we are able to apply this same decomposition of the Bernoulli for inference on the parameter of interest $\beta$. 

\subsection{A better way to do P2 fission for logistic regression}
\label{sec:our-logistic}

The problem in Example~\ref{ex_leiner_bernoulli} does not lie in  the Bernoulli P2 fission operation in Step 1 (see Example~\ref{ex:bernoulli}): instead, the problem lies in the fact that Step 3 uses a likelihood derived from the marginal distribution of $\yt$ rather than the conditional distribution of $\yt \mid \yo$. 

\begin{figure}[t]
\begin{floatrow}
\ffigbox{
\spacingset{1}
 \includegraphics[width=0.35\textwidth]{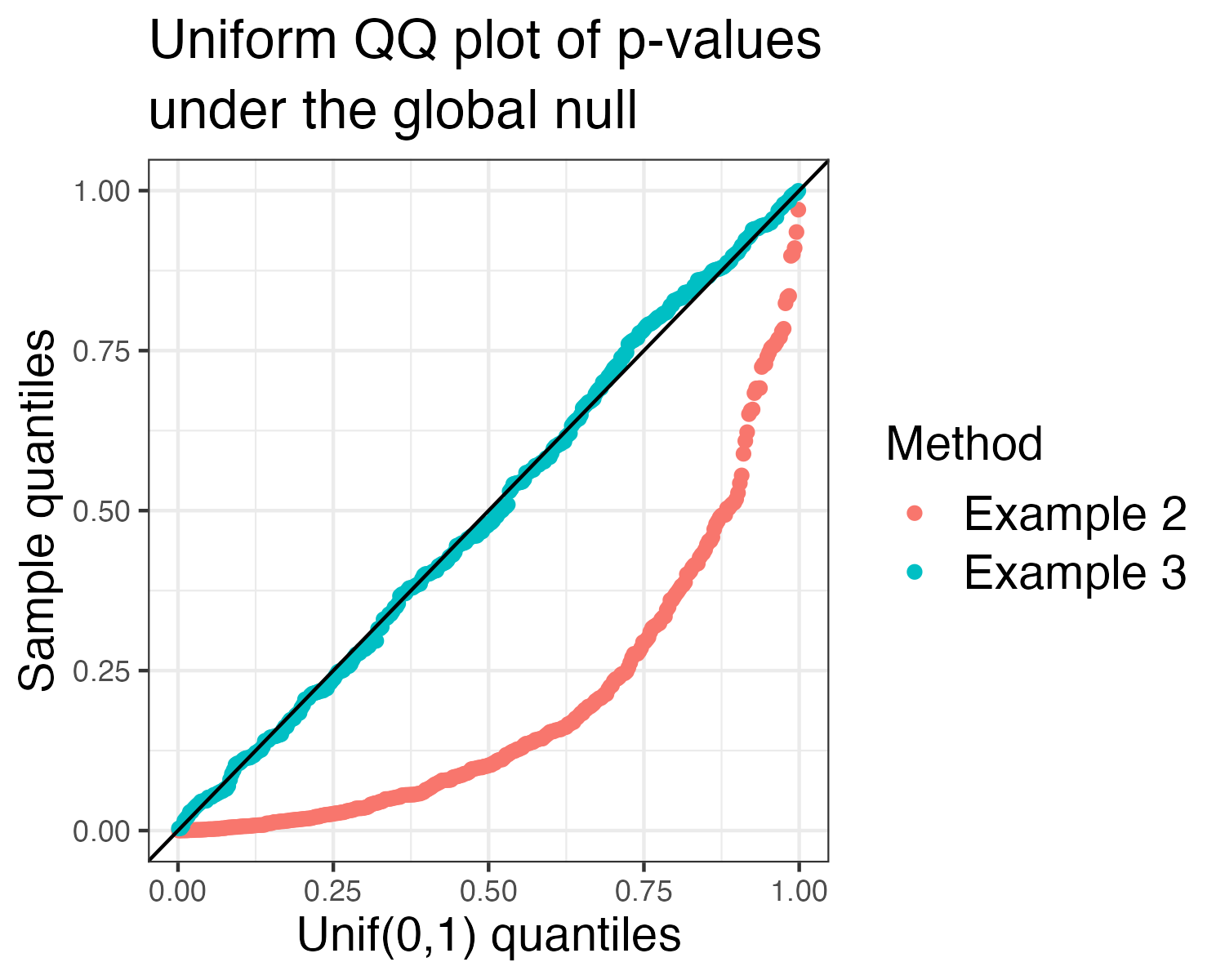}
}{
  \caption{{\em \footnotesize Uniform QQ-plot of p-values for all selected variables across all datasets under the global null, for the simulation described in Section~\ref{sec_logistic}. While the method of Example~\ref{ex_leiner_bernoulli} (taken from \cite{leiner2023data}) does not control the Type 1 error, our proposal in  Example~\ref{ex_leiner_bernoulli_corrected} achieves uniformly distributed p-values by directly working with the conditional distribution.}}
  \label{fig_globalnull}
}
\capbtabbox{%
\spacingset{1}
\scriptsize
\begin{tabular}{c|c|c|c}
$j$ & Coverage & Proportion of  & Proportion of datasets \\
& of $\beta_j$ & datasets w/  & w/ $X_j$ selected where \\
& & $X_j$ selected & $H_0: \beta_j=0$ is rejected \\
\hline
1 & 0.95 & 0.32 & 0.88 \\
\hline
2 & 0.94 & 0.94 & 1.00 \\
\hline
3 & 0.94 & 0.74 & 0.99 \\
\hline
Average over & 0.95 & 0.02 & 0.05 \\
all others 
\end{tabular}
\label{tab_mainresults}
}{%
  \caption{{\em \footnotesize Coverage, selection probability, and conditional power for the method in Example~\ref{ex_leiner_bernoulli_corrected}, for the simulation described in Section~\ref{sec:our-logistic}. The proportion of datasets with $X_j$ selected (for the non-null variables) is a function of how much information is in the training set, whereas the proportion of these datasets for which we reject the null hypothesis that $\beta_j=0$ is a function of the information in the test set. Thus, we can trade off the performance in these columns by changing the tuning parameter $\epsilon$ in Example~\ref{ex:bernoulli}: in these simulations, we use $\epsilon=0.8$.}}%
}
\end{floatrow}
\end{figure}

To address this, we re-consider  Step 3 of Example~\ref{ex_leiner_bernoulli}. 
It turns out that \eqref{eq_bernoulli_leiner_conditional} implies that 
\begin{align*}
\log \left( \frac{\mathrm{Pr} \left( Y^{(2)}  = 1 \mid Y^{(1)} \right)}{\mathrm{Pr} \left( Y^{(2)}  = 0 \mid Y^{(1)} \right)} \right)  &= \begin{cases}
\log\left(\frac{1-\epsilon}{\epsilon}\right)+ \beta_0 +\beta^T x_i & \mathrm { if } \ Y^{(1)}  = 1, \\ 
\log\left(\frac{\epsilon}{1-\epsilon} \right) + \beta_0+\beta^T x_i & \mathrm{ if } \ Y^{(1)} = 0. \\ 
\end{cases}
\end{align*}
Thus, we can improve on Step 3, as shown in the following example.
\begin{example}[Improved inference for logistic regression using P2 fission of a Bernoulli]\label{ex_leiner_bernoulli_corrected}
In the setting of Example~\ref{ex_leiner_bernoulli}, conduct Steps  1 and 2. Replace Step 3 with the following: 
\begin{enumerate}
\item[3.] Fit a standard logistic GLM to $\left\{ (x_{1S}, Y_1^{(2)}), \ldots, (x_{nS}, Y_n^{(2)}) \right\}$, with an offset that equals $\log\left(\frac{\epsilon}{1-\epsilon}\right)$ if $Y^{(1)}_i = 0$, and  $\log\left(\frac{1-\epsilon}{\epsilon}\right)$ otherwise. 
\end{enumerate}
    This method makes use of the conditional distribution of $\yt \mid \yo$, and therefore yields valid confidence intervals for the original parameters. 
\end{example}
Figure~\ref{fig_globalnull} shows that this leads to uniform p-values for the selected coefficients. 

We next generate $2,000$ datasets with $\beta_0=0.6$, $\beta_1=-0.9$, $\beta_2=2.1$, $\beta_3=-1.5$, and $\beta_4=\ldots=\beta_p = 0$. Table~\ref{tab_mainresults} shows that the standard (non-sandwich) logistic regression confidence intervals from Example~\ref{ex_leiner_bernoulli_corrected} achieve nominal coverage for each coefficient whenever it is selected. Table~\ref{tab_mainresults} also shows that the method from Example~\ref{ex_leiner_bernoulli_corrected} tends to select the truly important variables for the model, and has high power to determine that these variables are significant when they are selected. 

Thus, the conditional distribution of $\yt \mid \yo$ under the Bernoulli P2 fission operation in Example~\ref{ex:bernoulli} is quite tractable for logistic regression, and allows for valid post-selective inference using standard software. This provides a proof of concept  that P2 fission is a powerful tool that may be useful in practice. We believe that Example~\ref{ex_leiner_bernoulli_corrected}  makes this case more persuasively than the example in \citet{leiner2023data}'s supplement (re-stated as our Example~\ref{ex_leiner_bernoulli}), as in the latter example the parameters of interest are not the targets of inference. 
In the next section, we explore additional possible uses for P2 fission. 

Scripts to reproduce the numerical results in this section can be found at \url{https://github.com/anna-neufeld/fission_logistic/}. 

\section{P2 fission as P1 fission under misspecification}
\label{section_p2}

In this section, we argue that directly working with the conditional distribution of $\xt \mid \xo$ provides room for improvement in the way that previous authors have treated the setting of  P1 fission under model misspecification.

\subsection{Misspecified P1 fission of the Gaussian distribution}
\label{subsec_normal} 

We first recall the P1 fission operation for the Gaussian distribution with known variance from \cite{neufeld2024data}. This is equivalent (up to a constant scaling) to the operations of \cite{leiner2023data} and \cite{rasines2023splitting}.

\begin{example}[P1 fission of the Gaussian with known variance]
\label{ex_normal_unknown}
We observe $X \sim \mathrm{N}\left(\theta, \sigma^2\right)$. For a tuning parameter $\epsilon$, we let
\begin{equation}
\label{recipe_normal}
\left(\begin{matrix} \xo \\ \xt \end{matrix}\right)
\bigg\vert X=x \sim \mathrm{N}
\left( \left[ \begin{matrix} 
\epsilon x \\
(1-\epsilon) x 
\end{matrix} \right], \left[ \begin{matrix} \epsilon (1-\epsilon) \sigma^2 & - \epsilon (1-\epsilon) \sigma^2 \\
- \epsilon (1-\epsilon) \sigma^2 & \epsilon (1-\epsilon) \sigma^2 
\end{matrix} \right] \right) ,
\end{equation}
which yields  
\begin{equation}
\label{result_normal}
\left(\begin{matrix} \xo \\ \xt \end{matrix}\right) \sim \mathrm{N}\left( \left[ \begin{matrix} 
\epsilon \theta \\
(1-\epsilon) \theta
\end{matrix} \right], \left[ \begin{matrix} \epsilon \sigma^2 & 0 \\
0 & (1-\epsilon) \sigma^2 
\end{matrix} \right] \right).
\end{equation} 
\end{example}

If $\sigma^2$ is unknown, then we cannot draw from the distribution in \eqref{recipe_normal}. The question of what to do if $\sigma^2$ is unknown  has been considered by numerous authors. Both \cite{rasines2023splitting} and \cite{ leiner2023data} propose applying \eqref{recipe_normal} with an estimate $\hat{\sigma}^2$, and then treating the resulting folds $\xo$ and $\xt$ as though they were independent; both sets of authors justify this approach using an asymptotic argument. We refer to such an approach as ``approximate P1 fission".  However, since data fission is most useful when the number of observations $n$ is small (or even $n=1$), estimating $\sigma^2$ from the data is unlikely  to be fruitful in practical settings, and the asymptotic arguments of \cite{rasines2023splitting} and \cite{leiner2023data} are unlikely to apply. Thus, we feel that  ``approximate P1 fission"  leaves something to be desired.

Proposition 10 of \cite{neufeld2024data} provides the following finite sample result, which quantifies the magnitude of the correlation between $\xt$ and $\xo$ when we apply \eqref{recipe_normal} with a ``guess" of $\sigma^2$.  

\begin{example}[Gaussian P1 fission under misspecification]
\label{example_normal_p2}
Suppose that we observe $X \sim \mathrm{N}\left(\mu, \sigma^2\right)$, with both $\mu$ and $\sigma^2$  unknown. For a  $\tilde{\sigma}^2$ that is not a function of $X$, sampling
\begin{equation}
\label{recipe_normal_p2}
\left(\begin{matrix} \xo \\ \xt \end{matrix}\right)
\bigg\vert X=x \sim \mathrm{N}
\left( \left[ \begin{matrix} 
\epsilon x \\
(1-\epsilon) x 
\end{matrix} \right], \left[ \begin{matrix} \epsilon (1-\epsilon) \tilde{\sigma}^2 & - \epsilon (1-\epsilon) \tilde{\sigma}^2 \\
- \epsilon (1-\epsilon) \tilde{\sigma}^2 & \epsilon (1-\epsilon) \tilde{\sigma}^2 
\end{matrix} \right] \right)
\end{equation}
 yields
\begin{equation}
\label{result_normal_p2}
\left(\begin{matrix} \xo \\ \xt \end{matrix}\right) \sim \mathrm{N}\left( \left[ \begin{matrix} 
\epsilon \mu \\
(1-\epsilon) \mu
\end{matrix} \right], 
\left[ \begin{matrix} \epsilon^2 \sigma^2 + \epsilon (1-\epsilon) \tilde{\sigma}^2 & \epsilon (1-\epsilon) (\sigma^2 - \tilde{\sigma}^2) \\
\epsilon (1-\epsilon) (\sigma^2 - \tilde{\sigma}^2)  & (1-\epsilon)^2 \sigma^2 + \epsilon (1-\epsilon) \tilde{\sigma}^2 
\end{matrix} \right] \right).
\end{equation}
\end{example}
On the basis of this result, \cite{neufeld2024data} argue that  as long as $\sigma^2$ and $\tilde{\sigma}^2$ are ``close", the dependence between  folds can be ignored. 
Of course, in practice, it is not clear how an analyst might obtain an accurate ``guess" for the error variance, without using the data!

Notably, up to small changes in scaling and notation, Example~\ref{example_normal_p2} is also presented in  Section A of the supplementary materials of \cite{leiner2023data} as a Gaussian P2 fission operation for unknown ${\sigma}^2$. However, they do not elaborate on its use. 
%
In forthcoming work \citep{dharamshi2024decomposing}, we show that interpreting \eqref{result_normal_p2} as P2 fission provides a valuable toolset for  post-selection inference or model validation in the setting of a Gaussian with unknown variance, especially when $n$ is small or $n=1$. In particular, we work with the conditional distribution of $\xt \mid \xo$, rather than ignoring this dependence, as was done by previous authors \citep{leiner2023data,rasines2023splitting,neufeld2024data}. This conditional distribution poses a  number of statistical and computational challenges, which we address. We further explore topics such as the expected allocation of Fisher information between $\xo$ and $\xt \mid \xo$, which depends both on $\epsilon$ and the degree of misspecification. As in Section~\ref{sec:our-logistic}, we show that Gaussian P2 fission  enables inference on  parameters of interest using relatively standard techniques. 

As far as we know, \cite{dharamshi2024decomposing} are the first to propose using P2 fission as a remedy to P1 fission under misspecification. In the remainder of this section, we show that this idea can be fruitfully applied far beyond the Gaussian setting. 

\subsection{Misspecified P1 fission of the negative binomial distribution}
\label{subsec_nb}

We first restate the P1 fission operation for the negative binomial distribution with known overdispersion from \cite{neufeld2023negative} and \cite{neufeld2024data}.

\begin{example}[The negative binomial with known overdispersion]
\label{ex_nb_unknown}
We observe $X \sim \mathrm{NB}\left(r, \theta \right)$. If $r$ is known, then \cite{neufeld2024data} propose to sample $\xo$ and $\xt$ as
\begin{equation}
\label{recipe_nb}
\left(\xo, \xt \right) \bigg\vert X=x \sim \mathrm{DirichletMultinomial}\left(x, \epsilon r, (1-\epsilon) r \right),
\end{equation}
which yields $\xo \sim \mathrm{NB}\left(\epsilon r, \theta \right)$, $\xt \sim \mathrm{NB}\left((1-\epsilon) r, \theta \right)$, and  $\xo \perp\!\!\!\perp \xt$.
\end{example}

If the overdispersion parameter $r$ is unknown, then we cannot draw from the distribution in \eqref{recipe_nb}. \cite{neufeld2023negative} suggest that one can apply \eqref{recipe_nb} with an estimate of $r$ and treat the resulting folds as independent, and \cite{neufeld2024data} provide a finite sample result analogous to Example~\ref{example_normal_p2} that quantifies the covariance between the folds when a non-data-driven ``guess" $\tilde{r}$ is used. 

In the setting of Example~\ref{ex_nb_unknown}, 
suppose that we plug in a  very particular ``guess" $\tilde{r}$ in place of $r$:  we take $\tilde{r} \rightarrow \infty$. That is, we perform P1 fission for the incorrect family 
%
%
$\tilde{\mathcal{P}}$ 
$= \{ \mathrm{lim}_{r \rightarrow \infty} \mathrm{NB}\left(r, \theta \right) : \theta \in (0,1), r \frac{1-\theta}{\theta} = \mu \}$
$ =  \{ \mathrm{Poisson}\left( \mu \right) : \mu \in (0,\infty)\} $. 
\begin{example}[Applying Poisson P1 fission under misspecification] \label{ex:NB-poisson} We observe $X \sim NB(r, \theta)$, where both $r$ and $\theta$ are unknown. Apply Poisson P1 fission to $X$: that is, sample
$$
\left(\xo, \xt \right) \bigg\vert X=x \sim \mathrm{Multinomial}\left( x, \epsilon, (1-\epsilon) \right).
$$ 
This yields
$
\xo \sim \mathrm{NB}\left(r, \frac{\theta}{\theta+\epsilon-\epsilon \theta} \right)$ and $\xt \mid \xo \sim   \mathrm{NB}\left(r+\xo, \theta+\epsilon-\epsilon\theta \right)$.
This is identical to the P2 fission operation for the negative binomial given in  Appendix A of \cite{leiner2023data}. 
\end{example}
 Thus, the negative binomial P2 fission operation of \cite{leiner2023data} can be viewed as applying Poisson P1 fission to a negative binomial random variable.  It can therefore be interpreted as misspecified P1 fission.  

\cite{neufeld2024inference} and \cite{neufeld2023negative} explore the possibility of applying Poisson P1 fission to data that follows a negative binomial distribution, but they ignore the resulting dependence between folds of data, and suggest that this is appropriate if the amount of overdispersion in the data is thought to be mild. Negative binomial P2 fission offers an alternative path. However, the conditional distribution of $\xt \mid \xo$  takes a complicated form, which may pose a challenge in its application to post-selection inference or model validation. Understanding how to work with this conditional distribution in practice is a topic for future work. 

It is clear from Table~\ref{table_maintable} that there are many distributions beyond the Gaussian and the negative binomial for which the available P1 fission operation requires knowledge of some of the parameters of the model. Using the strategies presented in this section, we can see that it is always possible to come up with a P2 fission operation that does not require knowledge of any parameters: we can just take the P1 recipe and plug in a ``guess." This strategy will always result in dependent folds, and thus can always be seen as P2 fission. Thus, in this section, we have seen seen both a new way to deal with P1 fission under misspecification, as well as a possible future avenue for developing  new P2 fission decompositions.  For example, one could apply the P1 recipe for $\text{Beta}(\theta,\beta)$ referred to in Table~\ref{table_maintable} with a guess for $\beta$ to develop a valid P2 fission strategy for $\{\text{Beta}(\theta_1,\theta_2):\theta_1,\theta_2>0\}$, a family that does not yet appear in the table. 


\section{Discussion}
\label{sec_discussion}

We applaud \cite{leiner2023data} once again for their important work. 
 Decompositions of a single random variable have already proved useful in a variety of applications far beyond those mentioned in \cite{leiner2023data} \citep[see e.g.][]{tian2020prediction, chen2021estimating, oliveira2022unbiased, neufeld2024inference, neufeld2023negative}, and we have no doubt that the contributions of \cite{leiner2023data} will further increase the scope of application. 


In Section~\ref{sec_logistic}, we saw that in the setting of logistic regression,  operating on the conditional distribution of $\yt \mid \yo$ arising from P2 fission is actually quite tractable. This raises the following question: while many of the conditional distributions arising from the P2 fission operations proposed in \cite{leiner2023data} appear to be difficult to work with, might some of them be simpler than they seem? 
For instance, \cite{perry2024infer} consider a Gaussian-Laplace setting in which $\yt \mid \yo$ is complicated, but conditioning on additional information (beyond $\yo$) leads to tractable inference. Similar strategies may prove fruitful for other P2 fission decompositions.

Overall, while we would not choose P2 fission over P1 fission in a setting where both are available, we believe that P2 fission is a valuable tool whose potential is far broader than the examples considered in \cite{leiner2023data}. We look forward to seeing additional applications of P2 fission in the future. 

\section{Acknowledgments} We thank Daniel Kessler, Ethan Ancell, and Ronan Perry for helpful conversations that informed some of the ideas in this discussion. DW was partially supported by  ONR N00014-23-1-2589, NSF DMS 2322920, a Simons Investigator Award in Mathematical Modeling of Living Systems, and NIH 5P30DA048736. LG was supported by an NSERC Discovery Grant. AD was supported by an NSERC Postgraduate Scholarships-Doctoral. 




\bibliographystyle{plainnat}
\bibliography{fission}

\begin{thebibliography}{14}
\providecommand{\natexlab}[1]{#1}
\providecommand{\url}[1]{\texttt{#1}}
\expandafter\ifx\csname urlstyle\endcsname\relax
  \providecommand{\doi}[1]{doi: #1}\else
  \providecommand{\doi}{doi: \begingroup \urlstyle{rm}\Url}\fi

\bibitem[Chen et~al.(2021)Chen, Roch, Rohe, and Yu]{chen2021estimating}
Fan Chen, Sebastien Roch, Karl Rohe, and Shuqi Yu.
\newblock Estimating graph dimension with cross-validated eigenvalues.
\newblock \emph{arXiv preprint arXiv:2108.03336}, 2021.

\bibitem[Dharamshi et~al.(2024)Dharamshi, Neufeld, Motwani, Gao, Witten, and Bien]{dharamshi2024generalized}
Ameer Dharamshi, Anna Neufeld, Keshav Motwani, Lucy~L Gao, Daniela Witten, and Jacob Bien.
\newblock Generalized data thinning using sufficient statistics.
\newblock \emph{Journal of the American Statistical Association}, \penalty0 (just-accepted):\penalty0 1--26, 2024.

\bibitem[Dharamshi et~al.(2024+)Dharamshi, Neufeld, Gao, Bien, and Witten]{dharamshi2024decomposing}
Ameer Dharamshi, Anna Neufeld, Lucy Gao, Jacob Bien, and Daniela Witten.
\newblock {Decomposing Gaussians with unknown covariance}.
\newblock \emph{In preparation}, 2024+.

\bibitem[Joe(1996)]{joe1996time}
Harry Joe.
\newblock Time series models with univariate margins in the convolution-closed infinitely divisible class.
\newblock \emph{Journal of Applied Probability}, 33\penalty0 (3):\penalty0 664--677, 1996.

\bibitem[Leiner et~al.(2023)Leiner, Duan, Wasserman, and Ramdas]{leiner2023data}
James Leiner, Boyan Duan, Larry Wasserman, and Aaditya Ramdas.
\newblock Data fission: splitting a single data point.
\newblock \emph{Journal of the American Statistical Association}, pages 1--12, 2023.

\bibitem[Neufeld et~al.(2023)Neufeld, Popp, Gao, Battle, and Witten]{neufeld2023negative}
Anna Neufeld, Joshua Popp, Lucy~L Gao, Alexis Battle, and Daniela Witten.
\newblock {Negative binomial count splitting for single-cell RNA sequencing data}.
\newblock \emph{arXiv preprint arXiv:2307.12985}, 2023.

\bibitem[Neufeld et~al.(2024{\natexlab{a}})Neufeld, Dharamshi, Gao, and Witten]{neufeld2024data}
Anna Neufeld, Ameer Dharamshi, Lucy~L Gao, and Daniela Witten.
\newblock Data thinning for convolution-closed distributions.
\newblock \emph{Journal of Machine Learning Research}, 25\penalty0 (57):\penalty0 1--35, 2024{\natexlab{a}}.

\bibitem[Neufeld et~al.(2024{\natexlab{b}})Neufeld, Gao, Popp, Battle, and Witten]{neufeld2024inference}
Anna Neufeld, Lucy~L Gao, Joshua Popp, Alexis Battle, and Daniela Witten.
\newblock {Inference after latent variable estimation for single-cell RNA sequencing data}.
\newblock \emph{Biostatistics}, 25\penalty0 (1):\penalty0 270--287, 2024{\natexlab{b}}.

\bibitem[Oliveira et~al.(2022)Oliveira, Lei, and Tibshirani]{oliveira2022unbiased}
Natalia~L Oliveira, Jing Lei, and Ryan~J Tibshirani.
\newblock {Unbiased test error estimation in the Poisson means problem via coupled bootstrap techniques}.
\newblock \emph{arXiv preprint arXiv:2212.01943}, 2022.

\bibitem[Perry et~al.(2024)Perry, Xu, McGough, and Witten]{perry2024infer}
Ronan Perry, Zichun Xu, Olivia McGough, and Daniela Witten.
\newblock Infer-and-widen versus split-and-condition: two tales of selective inference.
\newblock \emph{arXiv preprint arXiv:2408.06323}, 2024.

\bibitem[Rasines and Young(2023)]{rasines2023splitting}
D~Garc{\'\i}a Rasines and G~Alastair Young.
\newblock Splitting strategies for post-selection inference.
\newblock \emph{Biometrika}, 110\penalty0 (3):\penalty0 597--614, 2023.

\bibitem[Sarkar and Stephens(2021)]{sarkar2021separating}
Abhishek Sarkar and Matthew Stephens.
\newblock {Separating measurement and expression models clarifies confusion in single-cell RNA sequencing analysis}.
\newblock \emph{Nature genetics}, 53\penalty0 (6):\penalty0 770--777, 2021.

\bibitem[Tian(2020)]{tian2020prediction}
Xiaoying Tian.
\newblock Prediction error after model search.
\newblock \emph{The Annals of Statistics}, 48\penalty0 (2):\penalty0 763--784, 2020.

\bibitem[Tian and Taylor(2018)]{tian2018selective}
Xiaoying Tian and Jonathan Taylor.
\newblock Selective inference with a randomized response.
\newblock \emph{The Annals of Statistics}, 46\penalty0 (2):\penalty0 679--710, 2018.

\end{thebibliography}

\appendix
\section{Proof of Proposition~\ref{prop1}}\label{app:info-proof}
As noted in Section 2.3 of \cite{leiner2023data},
$
I_X(\theta)=
I_{\tilde X^{(1)}}(\theta) + \mathbb E_{\tilde X^{(1)}}[I_{\tilde X^{(2)}\mid \tilde X^{(1)}}(\theta)].
$
For P1-fission, the same logic applies, but with $\xo$ and $\xt$ independent, this reduces to $I_X(\theta)=I_{\xo}(\theta)+I_{\xt}(\theta)$ (recovering Proposition 2 of \citet{dharamshi2024generalized}).
Setting these two decompositions of $I_X(\theta)$ equal to each other and recalling the assumption that the training sets have equal information, $I_{\xo}(\theta)=I_{\tilde X^{(1)}}(\theta)$, we get that
$$
\mathbb E_{\tilde X^{(1)}}[I_{\tilde X^{(2)}\mid \tilde X^{(1)}}(\theta)]=I_{\xt}(\theta).
$$
This equality together with Jensen's inequality then implies the result:
$$
\mathbb E_{\tilde X^{(1)}}[\{I_{\tilde X^{(2)}\mid \tilde X^{(1)}}(\theta)\}^{-1}]\ge \mathbb \{E_{\tilde X^{(1)}}[I_{\tilde X^{(2)}\mid \tilde X^{(1)}}(\theta)]\}^{-1}=I_{\xt}(\theta)^{-1}.
$$




\end{document}